\documentclass[prb,twocolumn,amsmath,amssymb]{revtex4}
\begin{document}
\author{Kevin Leung$^1$* and Martijn Marsman$^2$}
\affiliation{$^1$Sandia National Laboratories, MS 1415, Albuquerque, NM 87185,
{\tt kleung@sandia.gov}
\protect\\ $^2$Institut fuer Materialphysik, Universitat Wien,
Sensengasse 8/12, A-1090 Wien, Austria}
\date{\today}
\title{Energies of ions in water and nanopores within Density Functional
Theory}

\input epsf

\begin{abstract}
 
Accurate calculations of electrostatic potentials and treatment
of substrate polarizability are critical for predicting the
permeation of ions inside water-filled nanopores.  The {\it ab initio}
molecular dynamics method (AIMD), based on Density Functional
Theory (DFT), accounts for the polarizability of materials, water,
and solutes, and it should be the method of choice for
predicting accurate electrostatic energies of ions.  In practice, 
DFT coupled with the use of periodic boundary conditions
in a charged system leads to large energy shifts.
Results obtained using different DFT packages may vary because of the
way pseudopotentials and long-range electrostatics are implemented.
Using maximally localized Wannier functions, we apply robust corrections
that yield relatively unambiguous ion energies in select molecular
and aqueous systems and inside carbon nanotubes.  Large binding energies
are predicted for ions in metallic carbon nanotube arrays, while with
consistent definitions Na$^+$ and Cl$^-$ energies are found to exhibit
asymmetries comparable with those computed using
non-polarizable water force fields.  

\end{abstract}
 
\maketitle

\section{Introduction}
 
As computational power increases, Density Functional Theory
(DFT)-based aqueous phase
{\it ab initio} molecular dynamics (AIMD) simulations are becoming
more widely used. DFT accounts for the polarizability of water
molecules and can predict ion hydration structures significantly different
from those computed using non-polarizable force fields.  It also
accounts for the polarizability of synthetic nanopores
like carbon nanotubes\cite{benedict,marzari} and the surface
charge/dipole distribution-induced electrostatic potential inside
inorganic nanopores,\cite{pore,nenoff} which may strongly affect
ion permeation, rejection, partitioning, and exchange in these
water-filled systems.  
The ability to use AIMD to compare the intrinsic free energies of ions
in bulk water and inside water-filled nanopores may revolutionize our
understanding and modeling\cite{hummer_pnas,hummer_bio} of ion transport
in nanopores.  The predictions can potentially be tested using recently
synthesized membranes made of vertically aligned carbon
nanotube\cite{science,swnt_science} or functionalized silica
nanopores.\cite{brinker} But up to now, when charge states of simulation
cells are varied in the liquid phase, as when adding an ion
to a simulation cell, the anomalous average, absolute, ``intrinsic''
electrostatic potential $\phi_o$\cite{kleinman,deleeuw,redblack,saunders}
has hindered direct AIMD determination of energies.\cite{hunt,sprik2,sprik4}

In solid state DFT calculations, when Fourier
transform or Ewald techniques are used to compute long range electrostatic
interactions in periodically replicated unit cells, it has been widely
accepted that
$\phi_o$ is in general ambiguous.\cite{kleinman,deleeuw,redblack,saunders}
So is the energy incurred $q\phi_o$ when adding an ion of charge $q$ to a
charge neutral simulation cell.  Thus, successful estimates of
the energy difference between charged species in vacuum and those embedded
in a solid, such as the work function of electrons\cite{workfunc} or
in electrochemistrical applications,\cite{bell2} have generally
involved computing the {\it difference} in electrostatic potentials
across an explicit interface in a single simulation cell.  Since this
difference depends on the interface, which is not always
easily specified in complex systems, methods to reference the
intrinsic electrostatic potential in solids to some known values,
like the vacuum or Fermi level, have generated considerable
interest.\cite{saunders,bands,makov,schultz}
 
In contrast, in physical chemistry, techniques for using classical
force fields to compute the intrinsic free energy of ions in liquid water
have been well established and widely applied.\cite{preamble,pratt,pratt1,pratt2,pratt4,wood,darden,bell,levy,mccammon,ponder}
Here the electrostatic component of the ion hydration free energy is computed
using thermodynamic integration (TI) or free energy perturbation (FEP)
techniques.\cite{allen} It is added to the short-ranged repulsion and van
der Waal's contributions, which are often represented by Lennard-Jones terms.  

Formally, the predicted ion hydration free energies can be compared
with thermodynamic estimates of the absolute hydration free energies of
ions which are based on solvation and/or gas phase ion cluster
data.\cite{marcus,tiss}  To this end, the pure solvent contribution
to the water-vapor electrostatic ``surface
potential''\cite{pratt_sur2,pratt_sur,pratt_sur1,tildes1}
$\Delta \phi$ also needs to be computed.  Separating the calculation
into intrinsic and interfacial contributions is advantageous because
electrostatic effects are additive, and the pure solvent 
$\Delta \phi$ depends only on the water force field
used.\cite{pratt_sur1,tildes1}  Many widely used water
models yield $\Delta \phi$ on the order of $\sim 10$~kcal/mol
per proton charge.\cite{pratt_sur1,dang,tip5p}  At water-material
interfaces, $\Delta \phi$ will include the substrate contribution.
At vacuum-material interfaces, the substrate $\Delta \phi$ 
directly contributes to the work function and can be tested
against photoelectric measurements.\cite{workfunc,brocks} As will be shown,
incorporating parts of the substrate or pure solvent $\Delta \phi$
is crucial for a consistent definition of intrinsic DFT ion energies. 
This is our main interest in the $\Delta \phi$ term.

Note that there is also an ionic contribution to aqueous phase
$\Delta \phi$ which does not vanish even at infinite ion
dilution.\cite{friedman}  Further, it has been suggested $\Delta \phi$
at the water-vapor interface cannot be measured.\cite{pratt_sur2,guggen}
Nevertheless, the combination of intrinsic
ion free energy differences in bulk and confined water and the interfacial
$\Delta \phi$ may force ions into or reject them from narrow
nanopore entrances and lead to novel ion blockage behavior at finite
electrolyte concentration.  These effects will likely affect ionic
currents through nanoporous membranes.

\begin{figure}
\centerline{\hbox{\epsfxsize=3.20in \epsfbox{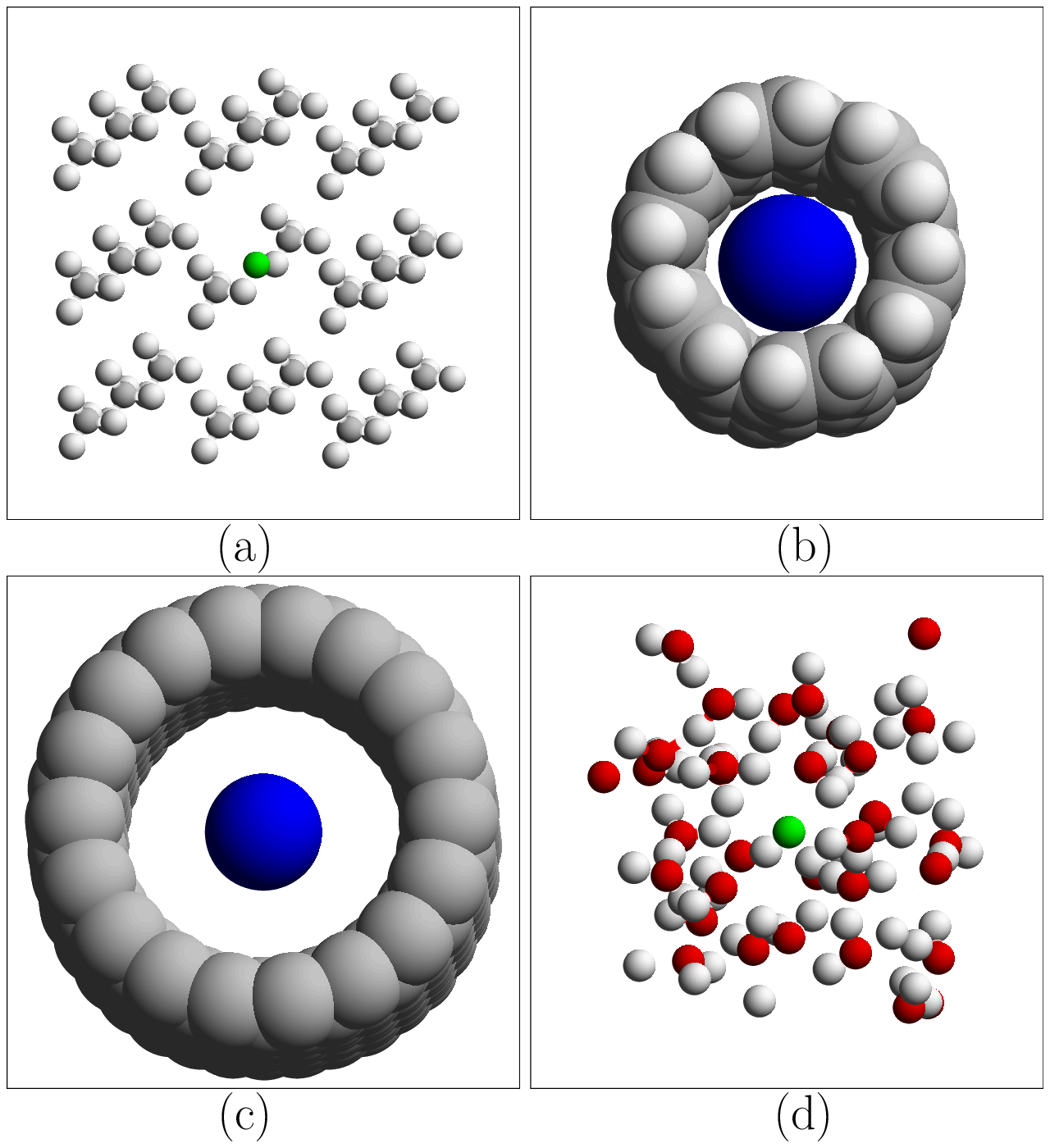}}}
\caption[]
{\label{fig1}}
Samples of the model systems examined in this work.  (a) CH$_4$ molecular
crystal; (b) (6,6) SWNT fragment; (c) infinitely long (18,18) SWNT;
(d) liquid water.  The Na$^+$ (Cl$^-$) ion is depicted in blue (green).
\end{figure}
                                                                                
In this work, we will use DFT to compute the intrinsic energy of ions
in some condensed phase systems without explicit reference to an
interface, just as has been done for ions in water using classical
force fields.  This requires regularizing the DFT
electrostatic potential, which will be performed in two different
environments: dipole-free molecular solids and metallic carbon nanotubes,
and bulk liquid water.  See Fig.~\ref{fig1}.  The philosophies differ, but
technically they both involve going beyond the well-known charge-image
and charge-neutralizing background correction\cite{makov,schultz,pratt}
and considering the ``spherical second moment'' electrostatic potential
contributions of the condensed phase material.\cite{saunders,makov,pratt_sur}
Unlike pure Ewald representation of electrostatics, our corrections
involve making physical choices, such as preserving the integrity of
molecules and nanotubes.  To facilitate this correction in liquid
water, the electronic charge density is decomposed on to molecular
centers using maximally localized Wannier
functions.\cite{wannier,berghold:prb:00}  To compare with DFT 
ion hydration results, we will focus exclusively on classical force
field simulations that also apply Ewald sums.
However, issues arising from truncating the coulomb
interaction\cite{pratt1,aqvist1,aqvist2} may also be relevant to
DFT calculations that do not use Ewald sums.

This paper is organized as follows.  Section~2 briefly describes the
computational methods.  It reviews how $\phi_o$ may be referenced to the
vacuum or a known interface value, and emphasizes and reconciles the
approaches taken in the solid state physics and physical chemistry
communities.  The results are presented in Sec.~3, and Sec.~4 concludes
the paper with further discussions.  An appendix describes the construction of
maximally localized Wannier functions within the projected augmented wave
(PAW) framework.\cite{paw0,paw}
 
\section{Method and Theory}

\subsection{Systems without dipole moments}

We consider 
\begin{equation}
E_{\rm ion} = E_{\rm ion + substrate} - E_{\rm bare~ion} - E_{\rm substrate}.
        \label{e_ion}
\end{equation}
In periodic boundary condition DFT calculations, $E_{\rm ion}$ contains
the generally unknown term $q \phi_o$, where $q$ is the charge of the
ion.  This ambiguity arises from the inherently long-ranged nature of
the Coulomb interaction $V({\bf r})=1/|{\bf r}|$, whose Fourier
representation $\hat{V}({\bf k})=4\pi/|{\bf k}|^2$ is undefined
at ${\bf k}=0$.  Ewald-sum calculations almost universally
and arbitarily set $\hat{V}({\bf k=0}) \propto \phi_o$ to zero.

In this work, we assume that any ion-containing periodically replicated
simulation cell has a neutralizing background, in effect removing the
monopole.  At zero temperature, the correction to $E_{\rm ion}$
due to the background and the image charge interactions has been well
documented in the DFT literature.\cite{makov}  It is
\begin{equation}
E_{\rm ewald}= q^2\alpha/(2L\epsilon) \label{ewald}
\end{equation}
in atomic units, where $\alpha$ is the Madelung constant, $L$ is the length
of the cubic unit cell, and $\epsilon = \epsilon_\infty$ or $\epsilon_o$
depending on the application.  When computing ion hydration free
energy in liquids using classical force fields, a similar correction
yield system size independent hydration free energies.\cite{pratt,pratt1,pratt2,pratt4,wood,darden,bell,levy,mccammon}
Equation~\ref{ewald} does not remove the
electric field arising from the periodic images, which is an area
of active research.\cite{schultz,martyna}  However, for the purpose of
this work, we will assume the monopole issue is a solved problem, and
will apply Eq.~\ref{ewald} and its generalization to non-cubic cells
with the appropriate $\epsilon$.  Instead, we focus on
the correction to the intrinsic electrostatic potential which
persists, and can be on the order of electron volts, even in the limit
of a very large unit cell where $E_{\rm ewald}$ becomes vanishingly small.
The origin of this shift in $\phi_o$ is discussed below
and is further illustrated in Sec.~\ref{crystal}.

In general, a simulation cell exhibits
finite dipole (${\bf d}$), quadrupole (the second rank tensor
${\bf {\tilde Q}}$), and ``spherical second'' (SSM, ${\bar Q}$) moments:
\begin{eqnarray}
{\bf d} &=& \int d{\bf r} ({\bf r}-{\bf R}_c) \rho({\bf r}); \label{dipole} \\
{\bf {\tilde Q}} &=& (1/2) \int d{\bf r} ({\bf r}-{\bf R}_c):({\bf r}-{\bf R}_c)
			 \rho({\bf r}); \label{second} \\
\bar{Q} &=& (1/3) {\rm Tr}{\bf {\tilde Q}} . \label{spherical}
\end{eqnarray}
Here ${\bf R}_c$ is an arbitarily chosen center of the unit cell,
$\rho({\bf r})$ is the total charge density, and the integration
over ${\bf r}$ in the unit cell includes both electronic and nuclear
degrees of freedom,
\begin{displaymath}
\int d{\bf r} \rho({\bf r}) \rightarrow \int d{\bf r}_{\rm e}
	\rho_e({\bf r}_{\rm e})
 + \sum_{m\beta} q_{m\beta} \delta ({\bf r}-{\bf R}_{m\beta}),
\end{displaymath}
where $\beta$ denotes the nuclear site on molecule $m$, and $e$ stands
for electrons.  

Unit cells with non-zero ${\bf d}$ and or ${\bf {\tilde Q}}$ elements
exhibit $\phi_o$ that depend on the surface terminations.  Consider
a macroscopic but finite object built from a periodically replicated
unit cell, terminated in interfaces with the vacuum.\cite{saunders}
If the lowest multiple moment of the the unit cell is a monopole,
dipole, or quadrupole, the center of the object (``crystal'')
experiences an electrostatic potential of the form
$\int^R dr r^2 f(r)$ exerted by the rest of the object.  Here
$|f(r)| \sim 1/r$, $1/r^2$, or $1/r^3$ respectively, and the vacuum region
outside the object, $r > R$, is assigned an electrostatic potential of
zero.  The integral is not uniformly convergent at large $R$, and thus
$\phi_o$ in unit cells at the center of the object depends on the shape
of the object in addition to the nature of the
interface.\cite{saunders} In particular, if
$|{\bf d}|\neq 0$, ${\bf {\tilde Q}}$ will depend on the choice
of ${\bf R}_c$, and there will be an ${\bf d}$-dependent electric field
across each unit cell.\cite{kleinman,deleeuw,redblack,saunders}

Ewald methods, when applied with the ``tin-foil'' boundary conditions (almost
universally the case in DFT calculations), effectively impose an infinite
$\epsilon_\infty$ dielectric continuum boundary that eliminates ${\bf d}$-
and ${\bf \tilde{Q}}$-induced electric fields and potentials.  As mentioned
above, it sets $\phi_o=\hat{V}({\bf k}=0)=0$ independent of
${\bf R}_c$.\cite{saunders}  To report $E_{\rm ion}$ in the
absence of an interface, the ``intrinsic'' non-zero $q\phi_o$, which
depends on physical choices of system boundaries, needs
to be added to the Ewald-computed ion energy.  Such a regularization of
DFT electrostatic potential within Ewald sum calculations has been carried
out in Ref.~\onlinecite{makov} for an isolated ion in the gas phase.  
The importance of this correction has not been sufficiently
emphasized in solid or liquid systems.

We first consider special cases where ${\bf d}$ vanishes or can be chosen
to vanish, and obvious physical choices of simulation cell boundaries
can be defined.  Such systems include ions inside a crystal made up
of molecules without permanent dipole moments, and ions embedded in
carbon nanotubes.  Consider the spherical second moment (SSM) contribution
in these systems.  This term does not appear in real space multipole
expansions of the electrostatic potential exerted by an arbitary charge
distribution.\cite{jackson}  But when Ewald sums arbitarily set $\phi_o=0$,
an SSM-dependent constant shift arises.  This shift is undone by
adding:\cite{saunders,makov}
\begin{equation}
\phi_{\rm SSM}({\bf R}_c) =
-(2 \pi /3 \Omega) \int d{\bf r} \rho ({\bf r})
	({\bf r}-{\bf R}_c)^2, \label{quad}
\end{equation}
where $\Omega$ is the unit cell volume.
If $\rho({\bf r})$ can be decomposed into molecular contributions labeled
by ``m'' and centered on molecular centers ${\bf R}_m$, such that
$\rho({\bf r})=\sum_m \rho_m({\bf r})$, then
\begin{eqnarray}
\phi_{\rm SSM}({\bf R}_{\rm c}) &=&
-(2 \pi /3 \Omega) \sum_m \bigg[ \int d{\bf r} \rho_m ({\bf r})
	({\bf r}-{\bf R}_m)^2 \nonumber \\
  &-2& \int d{\bf r} ({\bf R}_c-{\bf R}_m) \cdot ({\bf r}-{\bf R}_m)
		\rho_m({\bf r}) \bigg]  , \label{decom0}
\end{eqnarray}
where we have used the fact that each molecule has no net charge.
The position of the ion, if present, is assumed to coincide with
${\bf R}_c$, which permits the use of Eq.~\ref{decom0} without
adding monopole terms.
If each molecule also possesses a
zero dipole moment, the second term vanishes, and Eq.~\ref{decom0}
simplifies to
\begin{equation}
\phi_{\rm SSM}
 	= -(2 \pi /3 \Omega) \sum_m \int d{\bf r} \rho_m ({\bf r})
 	({\bf r}-{\bf R}_m)^2 \, . \label{decom}
\end{equation}
In this case, $\phi_{\rm SSM}$ no longer depends
on ${\bf R}_c$.  A system of identical, non-interacting molecules
without dipole (${\bf d}_m=0)$ or quadrupole moments,
($[{\bf {\tilde Q_m}}-1/3 {\rm Tr} {\bf {\tilde Q_m}}]_{ij}=0, i\neq j$), 
${\bf \tilde{Q}}_m = (1/2) \int d{\bf r} \rho_m({\bf r})
({\bf r}-{\bf R}_m):({\bf r}-{\bf R}_m)$,
would then yield a $\phi_{\rm SSM}$ proportional to the particle
density and the molecular-centered term
$\int d{\bf r} \rho_m({\bf r}) |{\bf r}-{\bf R}_m|^2$.
The absence of ${\bf d}_m$ means that ${\bf \tilde Q}_m$
is also independent of ${\bf R}_m$.

Several points are worthy of note.  (1) Unlike pure Ewald summation,
applying SSM corrections amounts to making physical choices of
system boundaries.  The electron density $\rho_e(r)$ is not uniquely
assigned to a simulation cell, but can be varied via ``spreading
transformations.''\cite{harris} For example, one can split
the $\rho_e(r)$ of a molecule which straddles the unit cell boundary,
creating two or more charge fragments; the $\phi_{\rm SSM}$ correction,
now no longer decomposable into intact molecular contributions,
will depend on how this splitting is performed.  Physically meaningful
interfaces consist of intact molecules without splitting of molecular
orbitals across unit cell boundaries.  When modeling ions in nanotubes,
the choice of boundaries will clearly affect the $q\phi_{\rm SSM}$
corrections to ion binding energies inside the pores.  Here the
simulation cells should preserve the $\rho_e(r)$ of intact SWNT in
the lateral directions.  Splitting of the continuous $\rho_e(r)$
along the axial direction is necessary, but intuitive choices
can still be made to demarcate $\rho_e(r)$ within each unit
cell of an infinitely long SWNT, guided by symmetry considerations.  In
covalently bonded solids like silicon, the choice of the unit cell is
less obvious.  (2) Note that carbon nanotubes are generally chemically
modified at their two
open ends, leading to surface contributions that may need to be added to the
computed intrinisic electrostatic potential when predicting ion transport
through synthetic nanoporous membranes.\cite{science,swnt_science,brinker}
Ionic contributions will also be present.\cite{friedman}
(3) $\phi_{\rm SSM}$ depends on whether core electrons or pseudopotentials
are used via Eq.~\ref{quad}.  Correcting $\phi_o$ with $\phi_{\rm SSM}$ 
is thus critical for comparing intrinsic ion energies computed
using different DFT codes and methods.  
(4) For unit cells without dipole moments, but with finite 
quadrupole moments, $[{\bf {\tilde Q}}-1/3 {\rm Tr} {\bf {\tilde Q}}]_{ii}
\neq 0$, $\phi_o$ still mathematically depends on the shape of the
macroscopic system, but not on unphysical considerations such as whether
core electrons are used.  (5) For isolated ionic species in vacuum,
$q\phi_{\rm SSM}$ vanishes when $\Omega \rightarrow \infty$.  In condensed
phase systems, where finite electronic and nuclear charge densities persist
in all space, $q\phi_{\rm SSM}$ becomes $\Omega$-independent for large
simulation cells, and is roughly proportional to the atomic density.
As will be seen, it can result in a several volt correction to the DFT-computed
$\phi_o$.  (6) As will be
discussed, classical force fields also exhibit $q\phi_{\rm SSM}$
corrections to the electrostatic potential, albeit smaller ones that
contain no ambiguity associated with ``core electrons'' contributions.

\subsection{Ions in liquid water}
\label{twice}

We also consider the energy of ions in bulk liquid water, where
Eq.~\ref{e_ion} is averaged over the AIMD trajectory.  Here $E_{\rm ion}$
is not the ion hydration enthalpy.  It does not account for the 
solvent reorganization penalty arising from molecular motion, and is
instead related to
the value of the integrand at the end point of the thermodynamic
integration formula where the ion charge $q_\lambda$ is gradually increased
from zero to $q$.\cite{pratt}  See Sec.~\ref{results_water}.
See Sec. III C.  However, regularizing $\phi_o$ 
is still crucial for correcting $E_{\rm ion}(q_\lambda)$ along the
entire $q_\lambda$-integration path.  As such, our work lays the
foundation for future AIMD thermodynamic integration calculations.
 
In general, a simulation cell containing an instantaneous
configuration of liquid water has non-vanishing dipole and quadrupole
moments.  Each water molecule also has finite ${\bf d}_m$ and
${\bf {\tilde Q}}_m$ elements.  The finite ${\bf d}_m$ means
that ${\bf {\tilde Q}}_m$ now depends on the choice of the molecular
center.\cite{jackson} Despite this, following formulations developed for
classical water force fields (see below), $q\phi_{\rm SSM}$
(Eq.~\ref{decom}) is still applicable because neither the vectors
$({\bf R}_c-{\bf R}_m)$ nor the molecular axes (implicit in
$\int d{\bf r}$) has preferred orientations, and the second term in
Eq.~\ref{decom0} averages to zero in liquid environments.  We
will collapse the electron density into molecular contributions via the
maximally localized Wannier function approach,~\cite{wannier,berghold:prb:00}
taking care to preserve the integrity of H$_2$O molecules and not cleaving
orbitals across unit cell boundaries.

Formally, comparing predicted absolute ion free energies with thermodynamic
estimates based on solvation and/or gas phase ion cluster data\cite{marcus,tiss}
requires adding the pure solvent contribution to $\Delta \phi$ at
the water-vapor interface (henceforth simply referred to as $\Delta \phi$
in this section).  The $\Delta \phi$ formulation is well established in
the classical force field literature:\cite{pratt_sur2,pratt_sur}
\begin{eqnarray}
\Delta \phi &=& \bigg\langle 4 \pi [-(1/3) \Delta \rho {\rm Tr}
{\bf {\tilde Q}}_m +\int_{z_1}^{z_2} \rho_d(z) ] \bigg\rangle
	\label{surface} \\
 &=& \Delta \phi_Q + \Delta \phi_d \nonumber \\
\rho_d(z) &=& \bigg\langle \int d{\bf r} \rho_m({\bf r})
\delta[z-({\bf R}_m)_z] ({\bf d}_m)_z \bigg\rangle \, . \label{surface1}
\end{eqnarray}
Here $z$ is the direction perpendicular to the interface, $\Delta \rho$
is the difference between water and vapor densities, and  $\rho_d(z)$
is the dipole density profile.  ${\bf d}_m$ and ${\bf {\tilde Q}}_m$
are the dipole vector and quadrupole moment tensor for water molecule $m$.
They are in fact independent of $m$ because of liquid
phase self-averaging (``$\langle ... \rangle$'').  Equation~\ref{surface}
would involve a sum over molecular species if more than one were present.
Its derivation makes use of the lack of preferred molecular orientations,
mentioned above.   Although
Refs.~\onlinecite{pratt_sur2}~and~\onlinecite{pratt_sur}, which
preceed Ref.~\onlinecite{saunders}, do not use the term ``second spherical
moment,'' the first term on the right side of Eq.~\ref{surface} is clearly
equivalent to $\phi_{\rm SSM}$ defined in Eq.~\ref{decom}.

Thus, the intrinsic ion hydration free energy computed using classical
force fields is typically defined without the $q\phi_{\rm SSM}$ term, which
is instead included as part of the pure solvent surface potential
$\Delta \phi$.\cite{hun_note,hun,hun1}
The numerical values of $\rho_d(z)$ and ${\bf {\tilde Q}}_m$ depend on
the choice of the molecular center ${\bf R}_m$, but
$E_{\rm ion}$+$q\Delta \phi$ (or more precisely the free energy
$\Delta G_{\rm ion}+q\Delta \phi$) is independent of such
choices.\cite{pratt_sur}

If we compute $q\Delta \phi$ using an AIMD simulation of the water-vapor
interface\cite{mundy} and add it to an appropriately defined AIMD intrinsic
ion hydration free energy, the result will be unambiguous.
However, using AIMD to compute $\Delta \phi$
is not yet feasible because (a) large numerical noises are associated with
the liquid-vapor interfacial fluctuations, and overly long and costly
AIMD trajectories would be required to yield good statistics for $\Delta \phi$.
(b) Widely used DFT functionals do not necessarily
yield 1.0~g/cc water density,\cite{mundy,dens} rendering the use of
Eq.~\ref{surface}, which depends on the difference between bulk
water and vapor densities $\Delta \rho$, problematic.

At present then, we believe AIMD predictions of ion free energies are most
useful for comparison with and validation of classical force fields.
This comparison requires that the $q\phi_{\rm SSM}$ correction be
consistently included as part of the intrinsic ion energy, and not
as part of $\Delta \phi$, for both AIMD and classical force field
simulations.  In fact, if they are both perfectly accurate and
consistently defined (using the same molecular centers ${\bf R}_m$),
DFT and classical force field calculations of $\Delta \phi_d$
(Eq.~\ref{surface1}), and not just the intrinsic ion energies,
should agree with each other.  This is because $\rho_d(z)$ only depends
on the spatial and orientational distributions of water molecules and
their instantaneous dipole moments at the water-vapor interface.  But
$\Delta \phi$ will not be the focus of this work; our interest in
this interfacial quantity is mainly to establish a consistent
definition of $E_{\rm ion}$ such that DFT, force field predictions,
and thermodynamic data\cite{marcus,tiss} can be compared.

Recall that $E_{\rm ion}$ neglects solvent reorganization cost arising
from molecular motion -- the dominant contribution to hydration in liquid
water.  Using Ewald-sums and non-polarizable classical force fields,
this definition of $E_{\rm ion}$ (or ``$q \phi$'' in the notation of Ref. 23)
becomes independent of the simulation cell
size when it is corrected by {\it twice} the value of
Eq.~\ref{ewald}.\cite{pratt}  We will apply this correction
to AIMD-predicted $E_{\rm ion}$ to compare with force field results.
Here the DFT electron density already ``reorganizes'' to accommodate the
ion and introduces a slight ambiguity.  The Ref.~\onlinecite{pratt}
correction will be shown to be reasonable by testing $E_{\rm ion}$
convergence with system size.  In contrast, the absolute ion hydration
free energy ($\Delta G$), a physical quantity, contains both electronic and
nuclear solvent reorganization.  It should be unambiguously cell-size
independent when Eq.~\ref{ewald} is subtracted {\it once}.\cite{pratt}
As mentioned above, for ions inside carbon nanotubes, electronic
relaxation dominates; we will freeze nuclear degrees
of freedom and show that Eq.~\ref{ewald} suffices for convergence.


To recapitulate, we consider two types of systems: ions in dipole-free unit
cells and in liquid water, both pertinent to ion penetration into
water-filled nanopores.  Different philosophies, traditionally associated
with solid and liquid phases, are involved, but the $\phi_{\rm SSM}$
corrections (Eqs.~\ref{quad}~and~\ref{decom}) are crucial in both
cases.  For a system of dipole- and quadrupole-free molecules, the two
viewpoints coincide.  $\phi_Q$ (Eq.~\ref{surface}) becomes
independent of the choice of molecular centers, $\rho_d(z)$ vanishes
identically, and Eq.~\ref{surface} reduces to Eq.~\ref{decom}.  In this case,
Eq.~\ref{surface} or Eq.~\ref{decom} directly references $E_{\rm ion}$
to $\phi_{\rm vacuum}=0$ through a material-vacuum interface that has no
dipole contribution to $\Delta \phi$ (Fig.~\ref{fig1}a).  

\subsection{VASP calculations}
 
We apply the Vienna Atomistic Simulation Package (VASP)\cite{paw,vasp}
and the standard VASP suite of ultrasoft (US)\cite{ultrasoft}
and projector-augmented wave (PAW) pseudopotentials (PP).\cite{paw0,paw}
US PP are used in AIMD simulations while static calculations utilize
PAW PP.  The Perdew-Wang (PW91) and Perdew-Burke-Ernzerhof (PBE) exchange
correlation functionals\cite{pw91,pbe} are applied for ions in bulk liquid
water and ions inside molecular solid/carbon nanotubes, respectively.  
These functionals generally yield similar predictions.
The Na$^+$ PP does not include pseudovalent $p$ electrons.
The AIMD time step is 0.25~fs, the cubic simulation
cells have sizes of 9.9874~\AA\, (12.509~\AA) for 32 (64) H$_2$O
molecules plus one ion, all proton masses are replaced by deuterium masses,
and the plane-wave energy cutoff is set at 400~eV.  The convergence criterion
at each of the Born Oppenheimer time steps is 10$^{-5}$~eV.
The total energy is conserved to within 1~K/ps (2~K/ps) for the Cl$^-$
(Na$^+$) simulation cells, respectively.  An elevated temperature of T=375~K
is enforced using a Nose thermostat to alleviate water-overstructuring
problems when applying the PW91 functional for AIMD
simulations.\cite{overstructure1,overstructure2,overstructure3}
The trajectory lengths used, not counting pre-equilibration runs, are at least
20~ps.  In one case (Cl$^-$ in 64 H$_2$O box), we actually use two different
initial configurations obtained from classical force field molecular dynamics
simulations, conduct two 10~ps AIMD trajectories, and combine the results.
The mean $E_{\rm ion}$ of the two AIMD segments are within 0.034~eV of each
other.  In the aqueous phase calculations, $E_{\rm ion}$ (Eq.~\ref{e_ion}) is
sampled every 0.1~ps.   $\Gamma$-point Brillouin zone sampling is applied
in most cases, except that a 1$\times$1$\times$10 $k$ Monkhorst-Pack grid is
also used in convergence tests for infinitely long metallic carbon nanotubes.

The implementation of the maximally localized Wannier functions
in VASP is limited to a $\Gamma$-point only sampling of the Brillouin zone.
Details concerning the construction of maximally localized Wannier
functions within VASP are provided in the Appendix.

To apply Eqs.~\ref{ewald} and~\ref{quad}, the VASP subroutine {\tt pot.F}
is slightly modified so that the electrostatic potential at
${\bf G}=0$, ${\bf k}=0$ reflects the total number of electrons (not
the negative of the total pseudo-nuclear charge).

\section{Results}
 
\subsection{Ions in molecular crystals}

\label{crystal}

We first consider two test cases of hypothetical simple cubic crystals
made up of Ar atoms and CH$_4$ molecules.  A lattice constant of 5~\AA\
ensures that the molecules are well-separated.  We construct a
3$\times$3$\times$3 supercell, create a vacancy at the supercell center,
and insert a Na$^+$ or Cl$^-$ ion there.  The Ewald corrections
(Eq.~\ref{ewald}) to eliminate the effects of the neutralizing background
cancel if we assume $\epsilon=1$ for both the isolated ion and the ion
inside the molecular crystal vacancy.\cite{note1} Centering the ion in
this supercell (``unit cell'') leads to zero net dipole and quadrupole
moments.  This physical choice of cell boundary allows unambiguous
computation of $q\phi_{\rm SSM}$ using Eq.~\ref{quad} (i.e., without
decomposing $\rho(r)$ into molecular
contributions).  Furthermore, for a macroscopic crystal made up of
this unit cell, any crystal-vacuum interface will not contribute to
$\phi_o$, $\Delta \phi$=0, and our predicted $E_{\rm ion}$ are
unambiguously referenced to the vacuum value $\phi=0$.

No geometric relaxation of the lattice atoms is allowed after inserting
Na$^+$ or Cl$^-$.  $E_{\rm bare~ion}$ for the isolated Cl$^-$
in a (15 \AA)$^3$ unit cell also exhibits a small $q\phi_{\rm SSM}= -0.1$~eV
using pseudopotential containing only 3$s$ and 3$p$ valence electrons. 
This is subtracted from the Cl$^-$-plus-substrate $q\phi_{\rm SSM}$ term.
With these corrections, we obtain $E_{\rm ion}=$-0.22 (-0.33)~eV for Na$^+$
(Cl$^-$) in the CH$_4$ solid, and $E_{\rm ion}=$-0.12 (-0.20)~eV in
Ar solid.  These values are reasonable.  With these large band gap
atoms/molecules, we expect small interactions between the ions and
the molecular dipoles they induce in the substrate.

\begin{table}\centering
\begin{tabular}{||c|c|c|c||} \hline
system & cell size & Na$^+$ & Cl$^-$ \\ \hline
CH$_4$(s) & 15.0$\times$15.0$\times$15.0 & -0.22 (-1.33) & -0.33 (+1.00) \\
Ar(s) & 15.0$\times$15.0$\times$15.0 & -0.12 (-1.66) & -0.20 (+1.48) \\
C$_{48}$H$_{24}$ & 15.0$\times$15.0$\times$15.0&-1.11 (-2.48) & -0.10 (+1.51) \\C$_{48}$H$_{24}$ & 20.0$\times$20.0$\times$20.0&-1.18 (-1.76) & -0.15 (+0.53) \\(6,6) & 14.8$\times$14.8$\times$14.8 & -2.85 (-6.29) & -1.77 (+2.14) \\
(6,6) & 14.8$\times$14.8$\times$22.2 & -2.86 (-6.26) & -1.80 (+2.16) \\
(18,18) & 17.0$\times$17.0$\times$17.0 & -2.20$^*$(-7.04) & -1.09$^\dagger$(+4.11) \\
                \hline
\end{tabular}
\caption[]
{\label{table1} \noindent
$E_{\rm ion}$ inside molecular crystals and SWNT's, in units of eV.
Simulation cell sizes are in \AA$^3$.  The values in brackets
are computed without the $q\phi_{\rm SSM}$ correction.  $^*$K$^+$
rather than Na$^+$ is used in the (18,18) SWNT; $^\dagger$
overestimated due to a small amount of electron transfer (see text).
}
\end{table}

More significantly, adding $q\phi_{\rm SSM}$ leads to comparable $E_{\rm ion}$
for both ions.  This is expected because the binding energy should
be roughly proportional to $q^2$.  In the absence of $q\phi_{\rm SSM}$
corrections, Na$^+$ and Cl$^-$ ions exhibit $E_{\rm ion}
\sim \pm 1$~eV in our model CH$_4$ and Ar solids
(see Table~\ref{table1}).  The large asymmetry between cations and anions,
and the predicted {\it repulsion} of Cl$^-$, mean that these uncorrected
DFT energies are clearly unphysical.  Table~\ref{table1} implies that
$\phi_{\rm SSM}>0$ in the DFT calculations.
This is because nuclei are extremely localized in space
whereas electrons are more diffuse.  The true electrostatic potential
$\phi_o$ averaged over the unit cell includes contributions inside atomic
cores where $\phi>0$.  Thus $\phi_o$ is positive definite inside the
crystal, {\it not} zero as Ewald sums typically mandate.  The position
independent $\phi_{\rm SSM}$ term correctly accounts for this.  Similar
corrections have been applied to non-polarizable Ne solid and Ar
fluid models using analytical atomic charge distributions.\cite{bands,hun}

The $\phi_{\rm SSM}$ correction is a consideration whenever classical force
fields with distributed charges and Ewald sums are used to compute intrinsic
hydration free energies of ions.\cite{sprik,cummings}  Even without
distributed charges, classical force field descriptions of molecules still
exhibit such corrections.  If the C and H atom sites in our CH$_4$ solid
carry fixed point charges of $-0.4e$ and $+0.1e$, $\phi_{\rm SSM}=-0.11$~volt.
Adding $q\phi_{\rm SSM}$ will then yield the expected physical
result, namely, that the ion energies in the vacancy site are almost zero
when computed using non-polarizable force fields.  Recall that
$\phi_{\rm SSM}$ can alternatively be defined as part of the interface
surface potential (Eq.~\ref{surface}) in force field calculations.

\subsection{Ions in carbon nanotube models}

Unlike the above molecular crystal test cases, the interactions of ions inside
single wall carbon nanotubes (SWNT) are of significant technological interest.
Narrow pore ($\sim 8$~\AA\, diameter) SWNT arrays have been proposed as osmotic
membranes,\cite{hummer_pnas} and axially aligned SWNT's with larger pore
diameters have recently been synthesized.\cite{science,swnt_science}  However,
most molecular dynamics simulations of water-filled
SWNT\cite{pore,hummer_pnas,hummer_bio,hummer_nature}
ignore SWNT polarizability,\cite{benedict,marzari} which is infinite for
metallic tubes, and large polarizability-induced effects are expected.

Figure~\ref{fig1}b depicts a proton-terminated (6,6) SWNT fragment
C$_{48}$H$_{24}$ situated in the center of a 15$\times$15$\times$15~\AA$^3$
cubic unit cell.  This finite-sized ``molecule'' has
a finite band gap and no dipole moments.
We place a Na$^+$ or a Cl$^-$ ion at the center of the C$_{48}$H$_{24}$
interior.  Ewald corrections (Eq.~\ref{ewald}) used to eliminate monopole
contributions are again assumed to cancel for the bare ions and the ion-SWNT
fragment complexes (see below).  The net $E_{\rm ion}$,
computed without relaxing SWNT atoms, are predicted to be -1.11~eV and
-0.10~eV for Na$^+$ and Cl$^-$,
respectively (Table~\ref{table1}).  Without $q\phi_{\rm SSM}$
(Eq.~\ref{quad}), they are erroneously predicted to be -2.6~eV and +1.6~eV.
Independent verification of this ion-position independent $q\phi_{\rm SSM}$
correction is obtained by moving the Na$^+$ from the tube interior
to an axial position halfway between the tube openings.  The energy difference
between these configurations is 1.17~eV.  Thus, $E_{\rm ion} \sim -1.11$~eV
indeed reflects the binding energy of Na$^+$ in C$_{48}$H$_{24}$.
We emphasize that, using non-polarizable force
fields,\cite{pore,hummer_pnas,hummer_bio} the binding energy would be purely
dispersive and would at most be a small fraction of an eV.

The binding energy should be independent of the simulation cell size
for this isolated SWNT fragment.  With a larger, $L=20$~\AA, cell
the $q\phi_{\rm SSM}$ corrected Na$^+$ (Cl$^-$) binding energy becomes 1.18~eV
(0.15 ~eV).  The small increase compared to $L=15$~\AA\, may
be due to the polarizability of C$_{48}$H$_{24}$ in the finite
simulation cell, which may make $\epsilon_\infty$ slightly larger than unity.
As $L\rightarrow \infty$, both $q\phi_{\rm SSM}$ and $(\epsilon_\infty-1)$
vanish as $1/L^3$.  Nevertheless, $|q\phi_{\rm SSM}|$ remains substantial,
$\sim 0.6$~eV, for the $L=20$~\AA\, cell (Table~\ref{table1}).  It
is much larger than that derived for isolated monoatomic
ions\cite{makov} because of the sheer number of electrons present in
C$_{48}$H$_{24}$.

Next, we consider Na$^+$/Cl$^-$ in infinitely long (6,6) and (18,18) SWNT
arrays.  These models mimic micron-thick carbon nanotube sieves which have
been synthesized.\cite{swnt_science}  The SWNT's chosen in
the present study are metallic and exhibit $\epsilon_\infty \rightarrow
\infty$ in the axial direction.  These nanotubes, with cubic unit
cell sizes chosen to reflect the SWNT periodicity
(14.8$\times$14.8$\times$14.8~\AA$^3$ and 17.0$\times$17.0$\times$17.0~\AA$^3$,
respectively), are not finite molecules.  However, by centering the ion
inside the SWNT and the SWNT within the simulation cell, 
and truncating the electron density at the $z$-direction of the grid boundary
--- taking care to evenly distribute $\rho(r)$ at the $z$-edges ---
we have made a physical choice of $\rho(r)$ in the unit cell that
exhibits no dipole moment and permits $q\phi_{\rm SSM}$ corrections.
The intrinsic energies of ions as they are displaced from the
pore center can then be referenced to the pore center value.

Equation~\ref{ewald} corrects the energies of the isolated ions.
Ions inside the SWNT are assumed to have $E_{\rm ewald}=0$ because of the
large $\epsilon_\infty$ of the SWNT array.  Table~\ref{table1} lists
the resulting $E_{\rm ion}$ for Na$^+$ and Cl$^-$.
Na$^+$ and Cl$^-$ exhibit large binding energies of 2.9 and 1.8~eV respectively
in the narrow (6,6) metallic SWNT arrays.  Increasing the Brillouin
zone sampling from $\Gamma$-point to a 1$\times$1$\times$10 Monkhorst-Pack
grid leads to changes in $E_{\rm ion}$ that are less than 0.02~eV.

Unlike the C$_{48}$H$_{24}$ fragment, the model SWNT arrays are meant
to be periodically replicated in the lateral directions.  We consider
system size dependences only by increasing the axial dimension of
the (6,6) simulation cell to 22.2~\AA.  We find that $E_{\rm ion}$
are conserved to within 0.03~eV after adding $q\phi_{\rm SSM}$, which are on
the order of several eV (Table~\ref{table1}).  This suggests that our
somewhat arbitary truncation of electronic density at the $z$-edges
nevertheless leads to robust $q\phi_{\rm SSM}$ corrections.
Unlike the Ar and CH$_4$ crystals, where $\Delta \phi=0$, $E_{\rm ion}$
cannot yet be referenced to vacuum values because we have not specified
the physical interface that will determine $\Delta \phi$.  However,
the two $q\Delta \phi$ should cancel when we consider
the sum of $E_{\rm ion}$ for the oppositely charge Na$^+$ and Cl$^-$.
Thus, regardless of the surface terminations, we predict substantial
binding energies for ions inside metallic SWNT.

No charge transfer occurs between Na$^+$ or Cl$^-$
and the (6,6) SWNT array.  DFT/PBE does predict that Na$^+$ accepts
a small fraction of an electron from the (18,18) array.  As a result,
we switch to the less electronegative K$^+$, whose lowest unoccupied
orbital is found to reside above the (18,18) Fermi level. Thus, K$^+$ is
representative of a monovalent cation well separated from the SWNT pore
surface.  It exhibits a still considerable 2.20~eV binding energy inside
the wider (18,18) array.  Cl$^-$ experiences a 1.09~eV binding energy.
This slightly overestimates the interaction of this SWNT with a
monvalent point anion because DFT/PBE predicts a small charge
transfer ($\sim 0.3$ electron) to the SWNT.

Studying ions inside SWNT is pertinent to permeation of
electrolytes, where the presence of water strongly stablizes
Na$^+$ and Cl$^+$ and helps prevent electron transfer to or from
the nanotubes.  How the strongly confined water will screen the induced
electrostatic interaction between ions and SWNT will be examined
in the future, but the SWNT polarizability clearly will have a
large impact on ion permeation, especially in the narrow (6,6)
SWNT that accommodates only a single file water
inside.\cite{hummer_pnas,hummer_nature}

\subsection{Ions in water}

\label{results_water}
 
\begin{figure}
\centerline{\hbox{\epsfxsize=3.20in \epsfbox{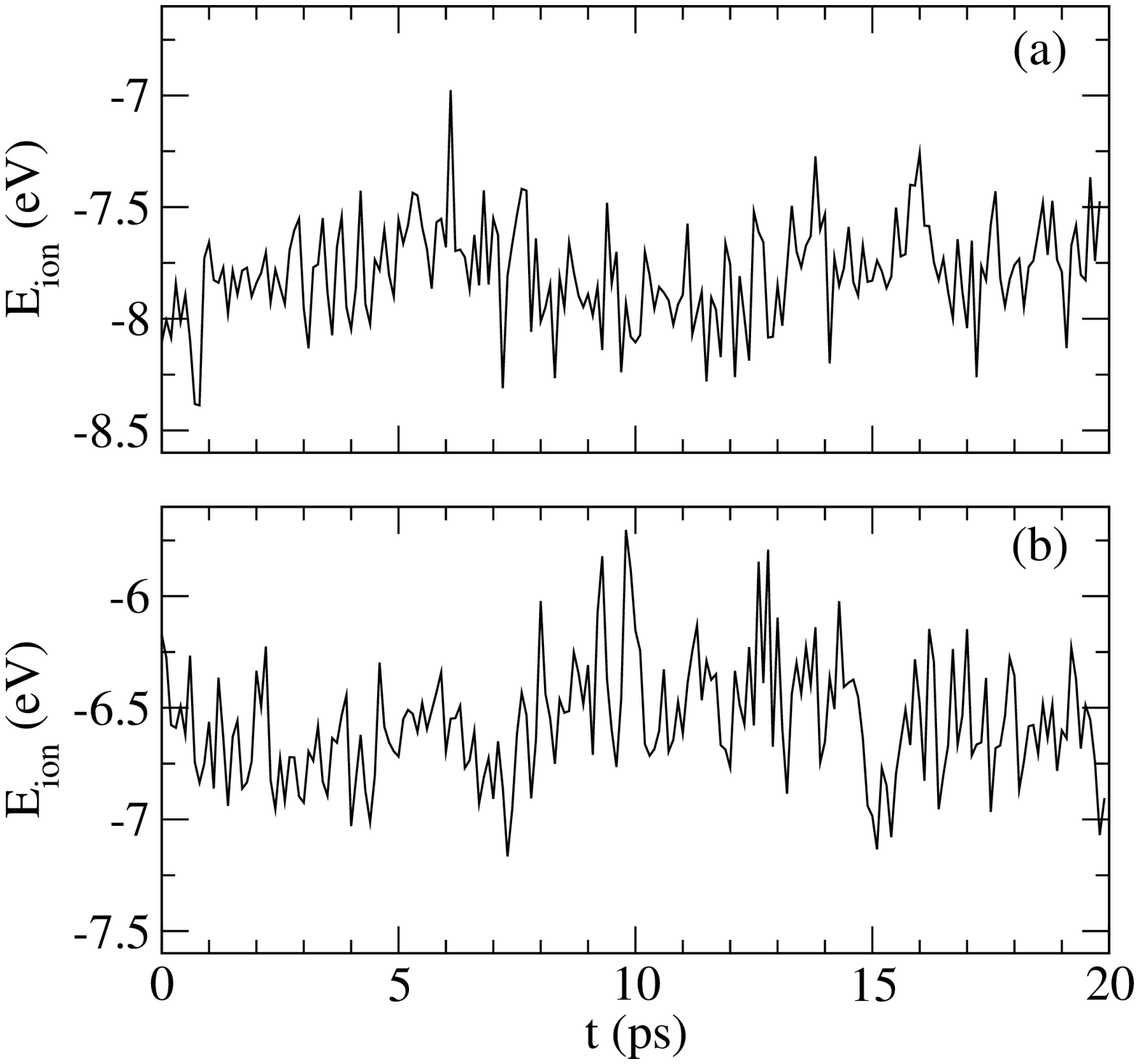}}}
\caption[]
{\label{fig2}}
$E_{\rm ion}$ as functions of time for (a) Na$^+$ and (b) Cl$^-$
in a simulation cell with 32 H$_2$O.  There is no significant drift
over these short 20~ps trajectories.
\end{figure}

In this section, we compute the mean $E_{\rm ion}$ for Na$^+$ and
Cl$^-$ in liquid water.  We apply the maximally localized Wannier
function method\cite{wannier,berghold:prb:00} to decompose electron
density on each H$_2$O.  This method correctly locates four Wannier
orbital centered within $\sim 1.5$~\AA\, of each water oxygen site,
corresponding to the two O-H covalent bonds and two lone pairs.  We
choose O as the molecular center in accordance with classical force
field conventions,\cite{pratt_sur} and add Ewald corrections
(twice that of Eq.~\ref{ewald}, as discussed above) and
$q\phi_{\rm SSM}$ to the instantaneous $E_{\rm ion}$ depicted in 
Fig.~\ref{fig2}.  $E_{\rm ion}$ exhibits little drift over these
20~ps trajectories.

Before we examine $E_{\rm ion}$ further, the robustness of using Wannier
functions to compute $q\phi_{\rm SSM}$ corrections (Eq.~\ref{decom})
needs to be documented.  Over a sample of 14 configurations, evenly
distributed over the Na$^+$ AIMD trajectory, we find that
$q\phi_{\rm SSM}$ experiences only $\sim 0.01$~eV variation at each
time step, amounting to a negligable standard deviation of 0.0028~eV.

\begin{figure}
\centerline{\hbox{\epsfxsize=3.20in \epsfbox{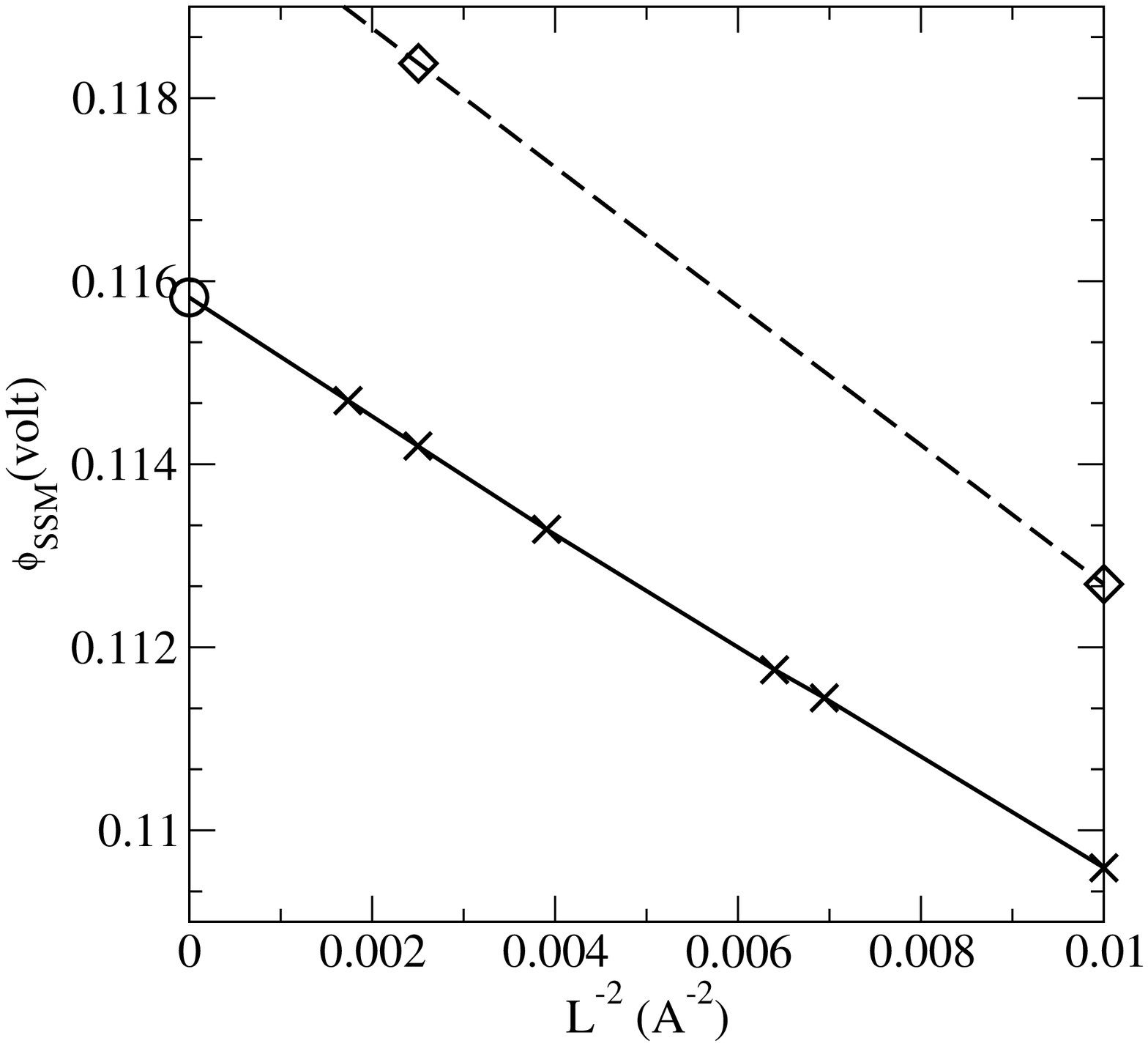}}}
\caption[]
{\label{fig3}}
$\phi_{\rm SSM}$ (Eq.~\ref{decom}, in units of volt per H$_2$O per
(10~\AA)$^3$) as the simulation cell size $L$ varies.  Crosses:
$\phi_{\rm SSM}$ for an isolated H$_2$O, using maximally localized
Wannier function.  Solid line: linear fit to scaling behavior.
Circle: $\phi_{\rm SSM}$ (Eq.~\ref{quad}) computed using
$\rho({\bf r})$ directly (without using Wannier functions).
Diamond and dashed line: $\phi_{\rm SSM}$ for 32~H$_2$O
($L=9.9874$~\AA) and 256~H$_2$O ($L=19.9748$~\AA).\cite{note2}
\end{figure}
                                                                                
Our maximally localized Wannier functions, by construction, depend somewhat
on the simulation cell size.~\cite{berghold:prb:00}
So do quantities we use to compute $\phi_{\rm SSM}$
within the Wannier function framework.
For a single water molecule in a cubic cell of length $L$,
the entire charge density belongs to the one H$_2$O, and the
Wannier function-derived Eq.~\ref{quad} can be compared with the value
obtained from numerical integration of $\rho_{\rm}(r)$.
Figure~\ref{fig3} shows that the Wannier-derived $\phi_{\rm SSM}$
convergence $\propto L^{-2}$;
$\phi_{\rm SSM}$ exhibits an empirical correction
$+0.623/L^2$ volt-\AA$^2$ per H$_2$O in a (10 \AA)$^3$ unit cell, or
$+623/L^5$ volt-\AA$^5$ per H$_2$O.  The scaling is consistent
with analysis of $\langle {\bf r} \rangle$ and $\langle r^2 \rangle$ 
described in the Appendix.  (Recall
$\phi_{\rm SSM}$ scales as 1/$\Omega$ for isolated molecules.)
Numerical integration over $\rho_{\rm}(r)$ yields $L$-independent
$\phi_{\rm SSM}$.
Its value agrees to within 0.01~\% of the Wannier $\phi_{\rm SSM}$
when the latter is extrapolated to $L\rightarrow \infty$.
This extrapolation increases $q\phi_{\rm SSM}$ computed in a
$L \sim 10$~\AA\, cubic cell by $\sim 5.5$~\%.  For water at 1.0~g/cc
density (32 H$_2$O molecules for this cell size), this increase
amounts to 0.22~volt.

Next, we show that the presence of multiple molecules does not substantially
change the $L$-dependence.  We expand a $L=9.9874$~\AA\, box of 32 water
molecules into a 2$\times$2$\times$2 supercell, and compare the
difference in $\phi_{\rm SSM}$ for this water configuration.  The
$L=19.9758$~\AA\ supercell $\phi_{\rm SSM}$ exceeds the $L=9.9874$~\AA\,
value by 5~\%.  According to the scaling curve in Fig.~\ref{fig3},
the Wannier function approach yields a 4.2~\% difference between
$L=20$ and $L=10$~\AA\, for an isolated H$_2$O.
This suggests that extrapolating $q\phi_{\rm SSM}$ computed for a
$L \sim 10$~\AA$^3$ box of liquid water to $L \rightarrow \infty$
at 1.0 g/cc density, using the $L$-scaling for a single H$_2$O,
should only lead to a small systematic error
of about (5-4.2)/5 $\times$ 0.22~volt, or about 0.035~volt.

\begin{table}\centering
\begin{tabular}{||c|c|c||} \hline
ion    & $N_{\rm w}$ &  $\langle E_{\rm ion} \rangle$ \\ \hline
Cl$^-$  &  32 & -6.57 (-2.60)  \\
Cl$^-$  &  64 & -6.27 (-2.29) \\
Cl$^-$$^\dagger$  &256 &  -7.39 (-8.23) \\
Na$^+$  &  32 & -7.85 (-11.54)  \\
Na$^+$  &  64 & -7.70 (-11.36) \\
Na$^+$$^\dagger$  &256 &  -10.01 (-9.17) \\ \hline
\end{tabular}
\caption[]
{\label{table2} \noindent
Mean $E_{\rm ion}$ in eV, accumulated using the
PW91 functional with (without) the $q\phi_{\rm SSM}$ correction.
$N_w$ is the number of water molecules in the simulation cell.
$^\dagger$SPC/E classical force field results (Table~4,
Ref.~\onlinecite{pratt}), which are independent of simulation cell size.
}
\end{table}

Having shown that $q\phi_{\rm SSM}$ corrections generated using maximally
Wannier functions are robust, and that the errors are well-controlled,
we return to $E_{\rm ion}$ in liquid water (Fig.~\ref{fig2}).
$E_{\rm ion}$ values averaged over at least 20~ps are listed in
Table~\ref{table2}.  $q\phi_{\rm SSM}$ is computed using Eq.~\ref{decom}
in the presence of the ion, with $q\phi_{\rm SSM}$ for the
isolated Cl$^-$ accounted for.  Removing the ion before computing
$q\phi_{\rm SSM}$ changes this term by less than 0.01~eV.  Overall,
$q\phi_{\rm SSM}$ is large --- on the order of several eV ---
for the water-filled simulation cell.  Na$^+$ and Cl$^-$
exhibit a 1.28~eV asymmetry in $E_{\rm ion}$.  To properly compare this
with force field results, we also apply the $q\phi_{\rm SSM}$ correction
to SPC/E water model used in Ref.~\cite{pratt} with oxygen as the molecular
center.  Because of the SPC/E
charge distribution, this {\it increases} ({\it decreases}) the magnitude
of $E_{\rm ion}$ for Na$^+$ (Cl$^-$).  These force field based Na$^+$
and Cl$^-$ $E_{\rm ion}$ now exhibit a 2.39~eV difference, which is
substantially larger than the DFT/PW91 result.  The
sum of the PW91 Cl$^-$ and Na$^+$ $E_{\rm ion}$, -14.42~eV, is also
substantially smaller than the -17.40~eV predicted using
the non-polarizable SPC/E water model.\cite{pratt}  The comparisons
depend on the force field used.

$E_{\rm ion}$ (Eq.~\ref{e_ion}) is the
end point of thermodynamic integration (TI) of ion hydration in systems
without electronic polarizability.  With polarizable water force fields,
it should be replaced by
$\langle d E_{\rm ion} /d (\lambda q_{\rm ion}) \rangle_{\lambda=1} $, where
$q_{\rm ion}$ is the charge of a non-polarizable ion force field.
AIMD treats water (and ion) polarizability, but this formulation is
in general not applicable because $q_{\rm ion}$ is set by
total number electrons in the system, and the highest occupied and lowest
unoccupied orbitals may not generally reside on the ion.  Instead, a
quantum mechanics/molecular mechanics (QM/MM)
approach with the {\it ion} treated classically, followed by free energy
perturbation to replace the MM ion by a DFT ion, may be a preferred
thermodynamic route to compute AIMD ion hydration free energies.

Thus, while
the ion hydration free energies reported in Ref.~\onlinecite{pratt}
are in reasonable agreement with experiments and the DFT results
are smaller and might suggest that DFT/PW91 underestimates monovalent
ion energetics in water, comparing the end-point integrand value of a
thermodynamic integration formula between non-polarizable force fields and
fully polarizable, DFT-based AIMD methods may be misleading.  Furthermore,
thermodynamic theories that incorporate DFT ion-water cluster energies
and dielectric continuum approaches have yielded good agreement with
experiments.\cite{rempe}  


Classical force field intrinsic $E_{\rm ion}$
exhibit $O(1)$~kcal/mol cell size dependences after 
correcting for image and background charge contributions.\cite{pratt}
However, there have been recent suggestions
that AIMD simulations exhibit larger cell size dependences for charged
species.\cite{sprik4}  Thus, we also consider a Na$^+$/Cl$^-$ plus
64 H$_2$O AIMD simulation at 1.0~g/cc H$_2$O density for 20~ps.
We find that $E_{\rm ion}$ changes by +0.15~eV (+0.30~eV) for Na$^+$
(Cl$^-$).  This amounts to a 3.5 (6.9)~kcal/mol $L$-dependence.

Are these $L$-dependences statistically significant?
It is difficult to estimate the uncertainty in $E_{\rm ion}$
using our short, $\sim 20$~ps AIMD trajectories (Fig.~\ref{fig2}).
Using the fact a 1~ns classical force field molecular dynamics
simulation typically yields a standard deviation of $\sim 0.5$~kcal/mol
in $E_{\rm ion}$, we estimate that our
20~ps AIMD simulations have 3~kcal/mol statistical uncertainties.  In
this sense, the $E_{\rm ion}$ difference between the 32- and 64-H$_2$O
cells is within the expected statistical errors.  
This test seems to suggest that our use of twice the value of Eq.~\ref{ewald},
$ 2\times 2.04$~eV, to correct the monopole-image interaction
leads to a reasonable system size convergence (see Sec.~\ref{twice}).
However, the AIMD solvent electronic polarizability also plays a role
in modifying the system size dependence seen in Ref.~\onlinecite{pratt}.
Further studies on ion hydration free energies are needed to address
this issue.  

\section{Conclusions}

In this paper, we apply DFT methods to compute the energies of ions
in condensed phase systems.  In general, due to the ambiguity in the average
intrinsic electrostatic potential $\phi_o$ arising from periodic boundary
conditions and Fourier transforms/Ewald sums, this energy is ill-defined.
It even depends on the number of electrons (valence vs.~core) and
computational protocol.  For two types of systems which are pertinent
to ion permeation in water-filled nanopores,  we demonstrate that
the intrinsic ion energy can nevertheless be reported.  The key 
is to apply the ``spherical second moment'' correction
($\phi_{\rm SSM}$),\cite{saunders,pratt_sur} in addition to the well-known
Ewald correction pertinent to a charged simulation cell with a
uniform neutralizing background.\cite{makov,pratt}  A physical choice
of unit cell boundary is involved in this correction.

The first type of systems consists of molecules or nanotubes whose unit
cell exhibits no permanent dipole moment. This enables relatively
unambiguous estimates of $q\phi_{\rm SSM}$, and yields ion energies that
can be referenced to the vacuum phase.  Note that the ion energy will
still mathematically depend on the shape of the crystal if the unit cell
contains a non-vanishing quadrupole moment.  Surface potential terms
may also contribute if the material surface is chemically different from
the bulk region.  Fortunately, interfacial effects are
additive to intrinsic ion energies.  With this approach, and choosing
cell boundaries that do not split up nanopore electronic densities in
radial directions, we have computed the interaction between an ion and model
molecular crystals and carbon nanotubes.  The intrinsic binding energies
for metallic (6,6) tubes are found to be 2.9 and 1.8~eV for Na$^+$
and Cl$^-$, respectively.  Even the much wider (18,18) SWNT's exhibit
binding energies of 2.2 and (a slightly overestimated) 1.1~eV for
monovalent cations and anions, respectively.  This suggests that
the proposed use of vertically aligned SWNT sieves as ion rejection
membranes\cite{pore,hummer_pnas,hummer_bio} may need to be re-examined.

The second type of systems consists of liquid-state simulation cells with no
obvious symmetry or other means to eliminate the net unit cell dipole
moment.  The philosophy is different here, and follow formulations derived
for classical water force fields, but $q\phi_{\rm SSM}$ corrections
are still useful because of orientational self-averaging in the liquid
state.  Here we apply maximally localized Wannier functions to decompose
charge densities onto molecules.  $\phi_{\rm SSM}$ will depend on the
choice of molecular center, which needs to be consistently defined
to compare with classical force field results.  It also involves
the physical choice that each water molecule is preserved intact
with no splitting of electron density across cell boundaries.  With consistent
$\phi_{\rm SSM}$ protocols and using a 32-water cell, we find that
Na$^+$ and Cl$^-$ exhibit ion energies (i.e., hydration energies
neglecting solvent reorganization) of -7.9 and -6.6~eV respectively.
These are considerably smaller than classical force field ion energies
of -9.0 and -7.1~eV.\cite{pratt}  Using a large simulation cell containing
64 H$_2$O molecules leads to 0.15 and 0.30~eV reductions in the ion energies.
These differences are within the statistical uncertainties of the simulations.
Note that our ``ion energy'' in the aqueous phase neglects solvent
reorganization and is not a measurable quantity.  It also differs from
the end-point of the thermodynamic integration formula for the hydration
free energy because of electronic polarizability issues.  Nevertheless,
with consistent $\phi_{\rm SSM}$ corrections, Na$^+$ and Cl$^-$ ion energies
become comparable, do not contain ambiguities associated core electrons
in pseudopotentials, and their asymmetry is now comparable to
force field predictions.  The ability to correct
for $\phi_o$ is important for future ion hydration free energy calculations
using AIMD in conjunction with thermodynamic integration methods.

Looking ahead, it is tantalizing to contemplate how the methods described
in this work can be applied to compute ion energies in crystalline
systems and nanoporous solids.\cite{nenoff}


\section*{Acknowledgement}
We thank Peter Schultz, Normand Modine, and Susan Rempe for useful
discussions and suggestions.  Sandia is a multiprogram laboratory
operated by Sandia Corporation, a Lockheed Martin Company, for the
U.S.~Department of Energy's National Nuclear Security Administration
under contract DE-AC04-94AL8500.
One of us (MM) was supported by the Austrian {\it Fonds zur F\"orderung
der wissenschaftlichen Forschung} within the START grant.

\appendix*
\section{\label{app:A}Maximally localized Wannier functions.}

The maximally localized Wannier functions $W_n$ are defined as linear
combinations of the $N_{\rm occ}$ selfconsistent Bloch wave functions
$\Psi_n$ that make up the space of occupied orbitals,
\begin{equation}\label{eq:wannier}
| W_n \rangle = \sum^{N_{\rm occ}}_{n'=1} U_{nn'} | \Psi_{n'} \rangle,
\end{equation}
where $U$ is the $N_{\rm occ} \times N_{\rm occ}$ unitary transformation
matrix that minimizes the {\it spread} functional,
\begin{equation}\label{eq:spread}
\Xi[U]=\sum^{N_{\rm occ}}_{n=1} \left( 
\langle W_n | r^2 | W_n \rangle - \langle W_n | {\bf r} | W_n \rangle^2
\right),
\end{equation}
which provides a measure for the degree of spatial localization of the
Wannier functions.
(Since the implementation of the maximally localized Wannier functions
in {\verb VASP } is limited to a $\Gamma$-point only sampling of the
Brillouin zone, we have dropped the usual ${\bf k}$-point index.)

Following Berghold {\it et al.}~\cite{berghold:prb:00} we define
\begin{equation}\label{eq:z}
z_{I,n}=\langle W_n | e^{i{\bf G}_I \cdot {\bf r}} | W_n \rangle,
\end{equation}
and rewrite Eq.~(\ref{eq:spread}) as
\begin{equation}
\Xi[U]=\frac{1}{(2\pi)^2}\sum^{N_{\rm occ}}_{n=1}\sum^{6}_{I=1}
\omega_I \left( 1- |z_{I,n}|^2 \right)+ O({\bf G}^2_I),
\end{equation}
where ${\bf G}_1=(1,0,0)$, ${\bf G}_2=(0,1,0)$, ${\bf G}_3=(0,0,1)$,
${\bf G}_4=(1,1,0)$, ${\bf G}_5=(1,0,1)$, and ${\bf G}_6=(0,1,1)$,
with respect to the basis of reciprocal lattice vectors.
The weights $\omega_I$ are defined as in Appendix A of
Ref.~\onlinecite{berghold:prb:00}.

In the PAW method\cite{paw0,paw} the one-electron
wave functions $\Psi_n$ are derived from the pseudo-wave-functions
$\widetilde{\Psi}_n$ by means of a linear transformation
\begin{equation}\label{eq:pawpsi}
|\Psi_n\rangle=|\widetilde{\Psi}_n\rangle+
\sum_{k} (|\phi_k\rangle-|\widetilde{\phi}_k\rangle)
\langle \widetilde{p}_k | \widetilde{\Psi}_n \rangle,
\end{equation}
The pseudo-wave-functions $\widetilde{\Psi}_n$ are the variational
quantities of the PAW method, and are expanded in reciprocal space
using plane waves.
The index $k$ is a shorthand for the atomic site ${\bf R}_k$, the angular
momentum quantum numbers $l_k$ and $m_k$, and the reference energy $\epsilon_k$.
The all-electron partial waves $\phi_k$ are solutions of the radial
Schr\"odinger equation for a non-spin-polarized reference atom at specific
energies $\epsilon_k$ and for a specific angular momentum $l_k$.
The pseudo-partial-waves $\widetilde{\phi}_k$ are equivalent to the
all-electron partial waves outside a core radius $r_c$ and match continuously
onto $\phi_k$ inside the core radius.
The projector functions $\widetilde{p}_k$ are dual to the pseudo-partial-waves
({\it i.e.}, $\langle \widetilde{p}_k | \widetilde{\phi}_l \rangle = \delta_{kl}$).

Following Ferretti {\it et al.},~\cite{ferretti:jpc:07} we use
Eqs.~(\ref{eq:wannier}) and (\ref{eq:pawpsi}), to rewrite
Eq.~(\ref{eq:z}) as
\begin{equation}\label{eq:zpaw}
\begin{split}
z_{I,n}=
\langle \widetilde{W}_n | e^{i{\bf G}_I \cdot {\bf r}}| \widetilde{W}_n \rangle +
\sum_{kl}
\langle \widetilde{W}_n | \widetilde{p}_k \rangle
\langle \widetilde{p}_l | \widetilde{W}_n \rangle \\
\times
\int
 e^{i{\bf G}_I \cdot {\bf r}}
\left[ \phi_k({\bf r}) \phi_l({\bf r})-
\widetilde{\phi}_k({\bf r})\widetilde{\phi}_l({\bf r})\right]
d{\bf r}.
\end{split}
\end{equation}
In our implementation, the integral over the partial waves in
Eq.~(\ref{eq:zpaw}) is approximated by
\begin{equation}
\sum_{L}\int_{\Omega_{{\bf R}_k}} \hat{Q}^{L}_{kl}({\bf r})
e^{i{\bf G}_I \cdot {\bf r}}d{\bf r},
\end{equation}
where $\hat{Q}^{L}_{kl}({\bf r})$ is given by Eq.~(27) of
Ref.~\onlinecite{paw}, and $\Omega_{{\bf R}_k}$ is the PAW
sphere at atomic site ${\bf R}_k$.
[Note that the double sum in Eq.~(\ref{eq:zpaw}) is limited to be
site-diagonal, {\it i.e.}, ${\bf R}_k={\bf R}_l$.]
An exact expression for the integral in Eq.~(\ref{eq:zpaw}) can be found
in App. B of Ref.~\onlinecite{ferretti:jpc:07}.

Thus defined within the PAW formalism, the spread functional
$\Xi[U]$ is minimized with respect to $U$ by means of the two-by-two
orbital rotation method introduced by Edminston and Ruedenberg
(see Sec. III A. of Ref.~\onlinecite{berghold:prb:00} and references therein).
Once the maximally localized Wannier functions are constructed,
we use
\begin{equation}
\label{eq:r2}
\langle r^2 \rangle_n =
\frac{1}{(2\pi)^2}\sum^{6}_{I=1}
\omega_I \left( 1- |z_{I,n}|^2 \right) +
\langle {\bf r} \rangle^2_n
+O({\bf G}_I^2),
\end{equation}
and
\begin{equation}
\label{eq:r}
\langle {\bf r} \rangle_n =
\langle W_n | {\bf r} | W_n \rangle =
\frac{1}{2\pi} \sum_{I=1}^{3} {\bf a}_I \Im \ln z_{I,n}
+O({\bf G}_I^2),
\end{equation}
where ${\bf a}_I$ are the real space lattice vectors, to
replace the electronic part of the integrals
$\int \rho_{m} ({\bf r}) {\bf r} d{\bf r}$ and
$\int \rho_{m} ({\bf r}) r^2 d{\bf r}$,
in Eqs.~(\ref{decom0}) and (\ref{decom}), by the sums
$\sum_i e \langle {\bf r} \rangle_i$ and
$\sum_i e \langle r^2 \rangle_i$
taken over the set of Wannier functions associated
with molecular center $m$.

Note that Eqs.~(\ref{eq:r2}) and (\ref{eq:r}) are exact to
order $O({\bf G}_I^2)$, which for simple cubic supercells
of lattice parameter $L$, means the expressions for
$\langle r^2 \rangle_n$ and $\langle {\bf r} \rangle_n$ converge
as $L^{-2}$ in the limit of large $L$ (see Fig.~\ref{fig3}).


\begin{references}
\bibitem{preamble}
In this work, we will focus on Ewald sum-based classical
force field calculations.
\bibitem{hun_note}
There have been recent controversies regarding whether the intrinsic ion
hydration free energies should be reported with or without the
$q\phi_{\rm SSM}$ correction.\cite{hun,hun1}  This seems to amount to
shifting $q\phi_{\rm SSM}$ from $q\Delta \phi$ (Eq.~\ref{surface})
to Eq.~\ref{e_ion}.  The free energy change of taking an ion from
vacuum, through the interface, into liquid, remains necessarily the
same.  However, this shift appears necessary for AIMD
to ensure that reported AIMD ion energies are independent of whether core
electrons are used.  Note that, when using Ewald sums, ``molecular''
or ``atomic'' based definitions of bulk water charge distributions\cite{hun}
yield identical electrostatic potentials.\cite{pratt2,wood}
\bibitem{note1}
In reality, $\epsilon_\infty$ should be slightly larger than unity
because of the polarizable molecules in the cell, even though they have
a large HOMO-LUMO gap and their density is low.  If the DFT
$\epsilon_\infty=1.05$, then according to Eq.~\ref{ewald}, $E_{\rm ion}$
for both ions would be underestimated by 0.06~eV.  The $\epsilon_\infty=1$
assumption should not alter the difference between these $E_{\rm ion}$.
\bibitem{note2}
At each cell size, the Eq.~\ref{decom}
correction is larger for in the condensed than the gas phase because
hydrogen bonding alters the charge distributions.
\end{references}
\end{document}